\documentclass[
reprint,
groupedaddress,
amsmath,amssymb,
aps,
prl
]{revtex4-2}

\usepackage{graphicx}
\usepackage{bm}

\begin{document}
	
	\title{Mechanism of State Transitions for the Vector Kuznetsov--Ma Breathers}
	
	\author{Muchen Dong}
	\author{Lei Wang}
	\email{Corresponding author: 50901924@ncepu.edu.cn}
	\affiliation{School of Mathematics and Physics, North China Electric Power University, Beijing 102206, People's Republic of China}
	
	\begin{abstract}
		We study two types of mechanisms of state transitions for vector Kuznetsov–Ma breathers (KMBs) in the coupled Fokas–Lenells framework on unequal backgrounds. The amplitude imbalance breaks
		spectral reflection symmetry and generates richer breather morphologies. In the degenerate sector, KMBs approaching a non-self-conjugate degeneration
		curve from opposite sides yield different limiting solitons, revealing
		a discontinuous KMB-to-soliton transition. In the nondegenerate sector, the background plane waves become frequency
		matched at a special wavenumber, converting the
		KMB into a single- or two-soliton state. We determine the exact transition threshold and characterize how nearby
		solutions vary with the parameter. We further investigate
		special limiting mechanisms arising on the self-conjugate degeneration curve
		and on the real spectrum at the state-transition wavenumber. Numerical excitation provides further evidence for these results.
	\end{abstract}
	
	\maketitle
	
	\paragraph{Introduction.---}
	Kuznetsov--Ma breathers (KMBs) are nonlinear excitations of a continuous-wave
	background: a localized structure periodically compresses and relaxes during
	its evolution \cite{Kuznetsov1977,Ma1979}. This recurrent exchange between the
	localized wave and its background connects breather dynamics with modulation
	instability and Fermi--Pasta--Ulam recurrence
	\cite{AkhmedievKorneev1986}. Such recurrence has been
	observed in modulationally unstable optical waves, and the full KMB evolution
	has been measured directly in optical fibers
	\cite{VanSimaeysEtAl2001,KiblerEtAl2012}. In multicomponent media,
	intercomponent coupling allows energy and intensity contrast to be
	redistributed between polarization or modal channels, so that the components
	can display inequivalent bright, dark, or multihump profiles. The Manakov
	system is the canonical integrable setting for this behavior, where
	nondegenerate vector KMBs exhibit branch-dependent morphologies, spectra, and
	soliton limits \cite{Manakov1974,CheEtAl2022}. Adding higher-order effects to
	the Manakov system leads to the coupled Hirota system, which exhibits a
	distinctive phenomenon known as a state transition: a breather is transformed
	into a soliton on specific parameter curves within particular regimes of
	modulational instability
	\cite{LiuEtAl2015StateTransition,PanEtAl2024Nondegenerate}. Such mechanisms
	are generally induced by higher-order effects, which are absent in the
	Manakov system.
	
	For ultrashort optical pulses, intrinsic space--time coupling and
	self-steepening can strongly reshape vector-wave dynamics
	\cite{BrabecKrausz1997}. The coupled Fokas--Lenells (CFL) system provides an
	integrable vector description of these effects
	\cite{Fokas1995,Lenells2009,ZhangEtAl2017}. Its operator \(D_\xi\) in the
	linear evolution term accounts for intrinsic space--time coupling, whereas
	its action in the nonlinear terms describes self-steepening. Because these
	derivative effects act on both the background waves and the localized core,
	they can control the relative phase of the two components and thereby
	reorganize the vector-wave morphology. The CFL system is already known to
	support bright--dark, dark--antidark, breather-like states and Peregrine
	solitons beyond the threefold limit
	\cite{ZhangEtAl2017,LingFengZhu2018,chenshihuaprl}. A critical question thus
	arises: do state transitions exist within the CFL system without higher-order
	terms?
	
	In this Letter, we show that amplitude imbalance breaks spectral reflection
	symmetry, enriches vector-KMB morphologies, and creates two distinct
	state-transition mechanisms. In the degenerate sector, approaching a
	non-self-conjugate degeneration curve from opposite sides produces two different
	dark--antidark limiting solitons, causing the limiting state
	to change discontinuously across the curve. In the nondegenerate sector, a
	special wavenumber matches the frequencies of the background plane waves and
	eliminates the relative temporal beating of the associated spectral modes.
	The KMB is then converted directly by the one-fold Darboux transformation
	into a single- or two-soliton state, without any limiting procedure, whereas
	the degenerate-sector solutions remain breathers. A pointwise
	total-intensity identity identifies this transition as a complementary
	redistribution of intensity between the two components, while the exact
	threshold shows that the secondary soliton disappears through escape and
	fading rather than merger. We further identify the previously known
	dark--antidark soliton family as the self-conjugate KMB limit and show that a
	real modulation component activates four localized waves in an otherwise
	inactive spectral interval. Direct numerical propagation supports these
	mechanisms.
	
	\paragraph{Model and spectral branches.---}
	For \(j=1,2\) and \(\ell=3-j\), let \(q_j(\xi,\tau)\) denote the \(j\)th field component, where \(\xi\) and
	\(\tau\) are the retarded time and propagation distance, respectively.
	The dimensionless CFL system reads
	\begin{equation}
		\label{eq:physical_cfl}
		{\rm i}D_\xi q_{j,\tau}-\tfrac{\eta}{2}q_{j,\xi\xi}
		+\left(2|q_j|^2+|q_\ell|^2\right)D_\xi q_j
		+q_jq_\ell^*D_\xi q_\ell=0.
	\end{equation}
	Here \(D_\xi=1+{\rm i}\nu\partial_\xi\), \(\eta=\pm1\) fixes the
	dispersion sign, and \(\nu>0\) measures the derivative correction. The transformations
	\(\xi=\nu\eta(x-t)\), \(\tau=-2\nu^2t\), and
	\(q_j={\rm i}e^{{\rm i}\eta(x+t)}u_j/(2\nu)\) give rise to
	\begin{equation}
		\label{eq:reduced_cfl}
		u_{j,xt}+u_j
		+{\rm i}\left(|u_j|^2+\tfrac12|u_\ell|^2\right)u_{j,x}
		+\tfrac{{\rm i}}2u_ju_\ell^*u_{\ell,x}=0.
	\end{equation}
	Since such transformations only rescale the amplitudes and linearly transforms
	the spatiotemporal plane, we henceforth study the reduced system
	\eqref{eq:reduced_cfl}.
	
	Consider the plane-wave background solutions
	\(u_{j,0}=a_j e^{{\rm i}(k_jx+\omega_jt)}\), where
	\(\omega_j=k_j^{-1}-a_j^2\), and set
	\(k_1=-k_2=k\), \(A=a_1^2\), \(B=a_2^2\),
	\(S=A+B\), and \(R=A-B\). The spatial eigenvalue \(\chi\) determines the squared Darboux parameter
	\(L=\lambda^2\) through
	\begin{equation}
		\label{eq:spectral_curve}
		L=\Lambda(\chi)=
		\frac{2(\chi^2-k^2)-Sk^2\chi+Rk^3}
		{\chi(\chi^2-k^2)},
	\end{equation}
	while the temporal eigenvalue is
	\(\Omega(\chi)=S/2+(2-kR)/(2\chi)\). Writing \(\chi_b=\chi_a+\gamma\) and
	\(\Lambda(\chi_a)=\Lambda(\chi_b)=L\) gives a quartic in \(\chi_a\).
	For a KMB, \(\gamma={\rm i}\xi\). The four roots naturally form two pairs,
	labeled by
	\(\operatorname{Re}\chi_{1,2}<Sk^2/4<
	\operatorname{Re}\chi_{3,4}\). The roots within each pair generate symmetry-related and morphologically
	equivalent profiles, whereas the two pairs generally yield inequivalent vector states, termed the nondegenerate vector KMBs.
	
	Define \(h_1(\chi)=k/(k+\chi)\), \(h_2(\chi)=k/(k-\chi)\),
	\(\Omega_\alpha=\Omega(\chi_\alpha)\) for \(\alpha=a,b\), and
	\(
	\mathcal E=\rho_0
	e^{{\rm i}[(\chi_b-\chi_a)x+(\Omega_b-\Omega_a)t]},
	\)
	where \(\rho_0\neq0\) is the relative modal weight. Furthermore, let
	\(H_j=h_j(\chi_a)+h_j(\chi_b)\mathcal E\) and
	\(\mathcal Q=A|H_1|^2+B|H_2|^2\), one has
	\begin{equation}
		\label{eq:compact_dt}
		u_j^{[1]}=u_{j,0}\left[1+
		\frac{(L^*-L)H_j(1+\mathcal E^*)}
		{L|1+\mathcal E|^2-\mathcal Q}\right],\quad j=1,2.
	\end{equation}
	Each KMB is determined by the root pair \((\chi_a,\chi_b)\) and modal
	weight \(\rho_0\). For unequal amplitudes (\(R\neq0\)), the cubic term
	\(Rk^3\) is nonzero and breaks the \(k\mapsto-k\) symmetry, making both
	the KMB morphology map and its existence domain asymmetric in \(k\), and
	thereby enabling state-transition mechanisms absent in the commonly
	studied equal-amplitude setting.
	
\begin{figure}[h]
		\centering
		\includegraphics[width=0.5\textwidth]{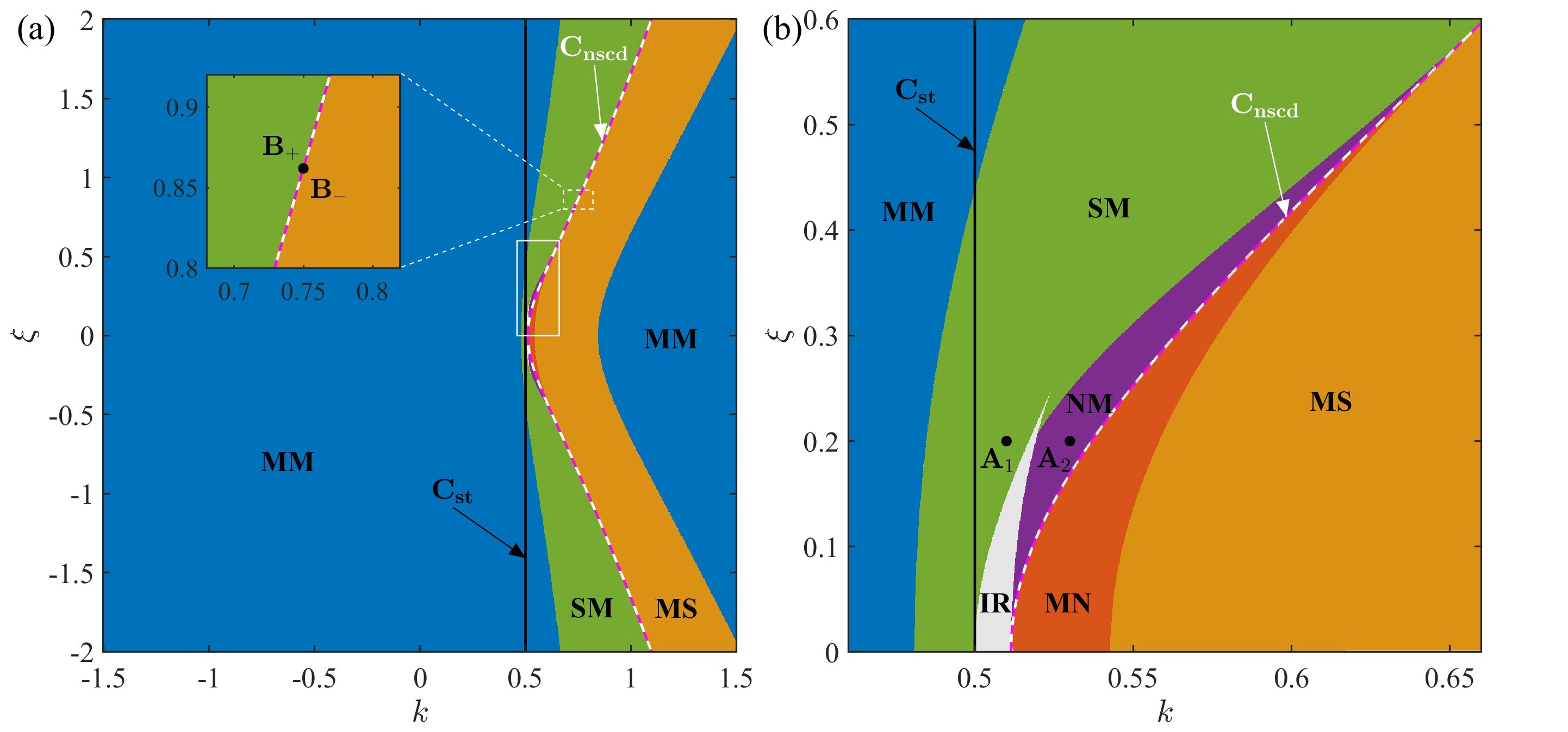}\vspace{-1em}
		\caption{KMB morphology map for
			\(\chi_a\in\{\chi_1,\chi_2\}\) at \(A=5,B=1\):
			(a) global view and (b) enlargement of the white box in (a).
			Two-letter labels give components 1 and 2, with \(M,N,S\) denoting
			bright, dark, and four-petal structures (e.g., \(MM\) is
			bright--bright); \(IR\) denotes the invalid region. \(C_{\rm st}\) and
			\(C_{\rm nscd}\) are the state-transition and non-self-conjugate
			degeneration curves. \(A_{1,2}\) give Figs.~\ref{fig:degenerate_profiles}(a,b),
			and the one-sided limits at \(B_{\pm}\) give
			Figs.~\ref{fig:nonself_boundary}(a,b).}\vspace{-1em}
		\label{fig:existence}
	\end{figure}

\begin{figure}[h]
		\centering
		\includegraphics[width=0.4\textwidth]{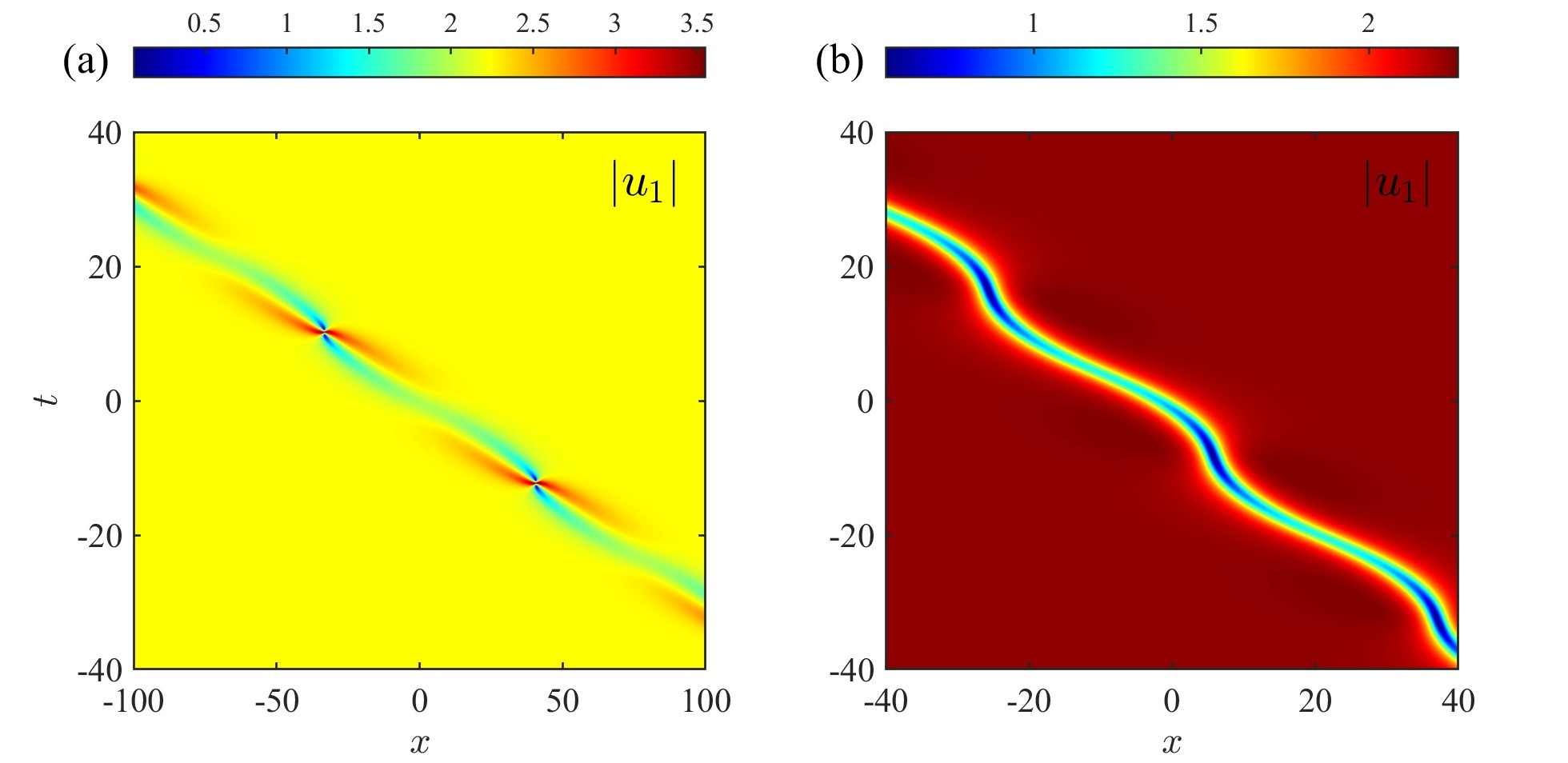}\vspace{-1em}
		\caption{First-component profiles of KMBs for
			\(A=5,B=1\): (a) the four-petal KMB at
			\((k,\xi)=(0.51,0.20)\); (b) the dark KMB at
			\((k,\xi)=(0.53,0.20)\).}\vspace{-2em}
		\label{fig:degenerate_profiles}
	\end{figure}

	\paragraph{Degenerate KMB region and non-self-conjugate degeneration.---}
	\textbf{Fig.~\ref{fig:existence}} maps the KMB morphologies obtained by choosing
	\(\chi_a=\chi_1\) or \(\chi_2\). Within this sector, amplitude imbalance
	enables a four-petal first-component KMB at \(A_1\) and a dark one at \(A_2\),
	as shown in \textbf{Fig.~\ref{fig:degenerate_profiles}}. Neither morphology occurs on
	the equal-amplitude background, where the first component remains bright.
	
	
\begin{figure*}[t]
		\centering\vspace{-1em}
		\includegraphics[width=0.8\textwidth]{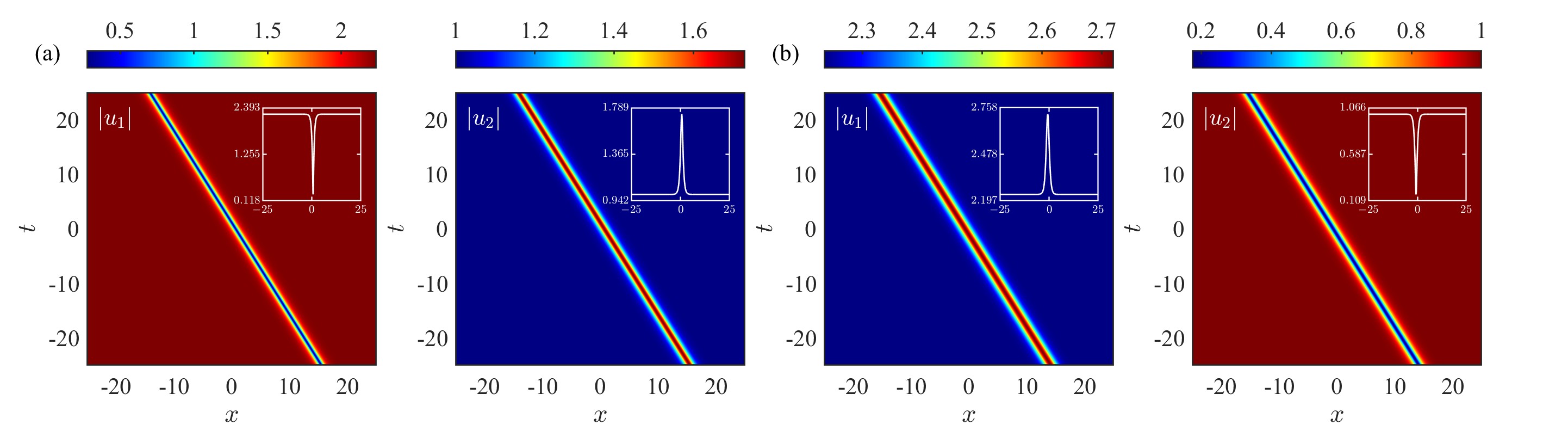}\vspace{-1em}
		\caption{One-sided limit solitons on \(C_{\rm nscd}\) for \(A=5,B=1\) at
			\((k,\xi)=(0.75,0.8619)\):
			(a) dark--antidark soliton at \(B_+\);
			(b) antidark--dark soliton at \(B_-\).}\vspace{-2em}
		\label{fig:nonself_boundary}
	\end{figure*}

	Within the degenerate region, where the choices
	\(\chi_a=\chi_3,\chi_4\) do not generate admissible KMBs, there exists a
	non-self-conjugate degeneration curve \(C_{\rm nscd}\),
	\begin{equation}
		\label{eq:F_nonself}
		2\bigl(9Rk^3-16k^2-4X_{\rm ns}\bigr)^2
		+18S^2k^4X_{\rm ns}
		-27S^2k^6(Rk-2)=0,
	\end{equation}
	where \(X_{\rm ns}=\xi_{\rm ns}^2\),
	\(\alpha=(9Rk^3-16k^2-4X_{\rm ns})/(3Sk^2)\), and
	\(\mu=2/(3\alpha)\). On this curve, the three distinct roots
	\(\chi_a=\alpha-{\rm i}\xi_{\rm ns}\), \(\chi_b=\alpha\), and
	\(\chi_c=\chi_a^*=\alpha+{\rm i}\xi_{\rm ns}\) share the same real
	spectral value,
	\(\Lambda(\chi_a)=\Lambda(\chi_b)=\Lambda(\chi_c)=\mu\in\mathbb R\).
	Consequently, \(L^*-L=0\) and the Darboux solution reduces to
	the background plane wave. However, nontrivial limits survive when \(C_{\rm nscd}\) is approached
	through complex spectral values. For the two orientations
	\(\chi_E=\alpha+{\rm i}\varepsilon\xi_{\rm ns}\) and
	\(\chi_0=\alpha\), with \(\varepsilon=\pm1\), let
	\(L=\mu+{\rm i}\epsilon_L\) and impose
	\(\mathcal E=\sqrt{\epsilon_L}\,e^{-\Theta_\varepsilon}\) so that the
	Darboux numerator and denominator vanish at the same order. The resulting limiting solitons are
	\begin{equation}
		\label{eq:nonself_limit}
		u_j^{{\rm ns},\varepsilon}
		=u_{j,0}\left[
		1-\frac{2{\rm i}h_j(\chi_E)}
		{d_\varepsilon-\Delta_\varepsilon
			e^{-2\Theta_\varepsilon}}\right],
		\qquad j=1,2,
	\end{equation}
	where
	\[
	\Theta_\varepsilon
	=-\varepsilon\xi_{\rm ns}x
	+\operatorname{Im}\!\left[
	\Omega(\chi_0)-\Omega(\chi_E)\right]t
	-\ln\rho_\varepsilon
	\]
	is the comoving coordinate, with \(\rho_\varepsilon>0\) fixing the
	soliton position. The finite spectral coefficients \(d_\varepsilon\) and
	\(\Delta_\varepsilon\), complex and real respectively, are determined by
	\(A,B,k,\alpha,\xi_{\rm ns}\), and the orientation \(\varepsilon\). As \(C_{\rm nscd}\) is approached from its two sides, represented by
	\(B_+\) and \(B_-\), the KMB evolves continuously into dark--antidark and
	antidark--dark limiting solitons, respectively, as shown in
	\textbf{Fig.~\ref{fig:nonself_boundary}}. To our knowledge, this discontinuous
	limiting transition has not been reported previously and constitutes the
	first state transition identified in a degenerate KMB region.
	
\begin{figure}[h]
		\centering
		\includegraphics[width=0.5\textwidth]{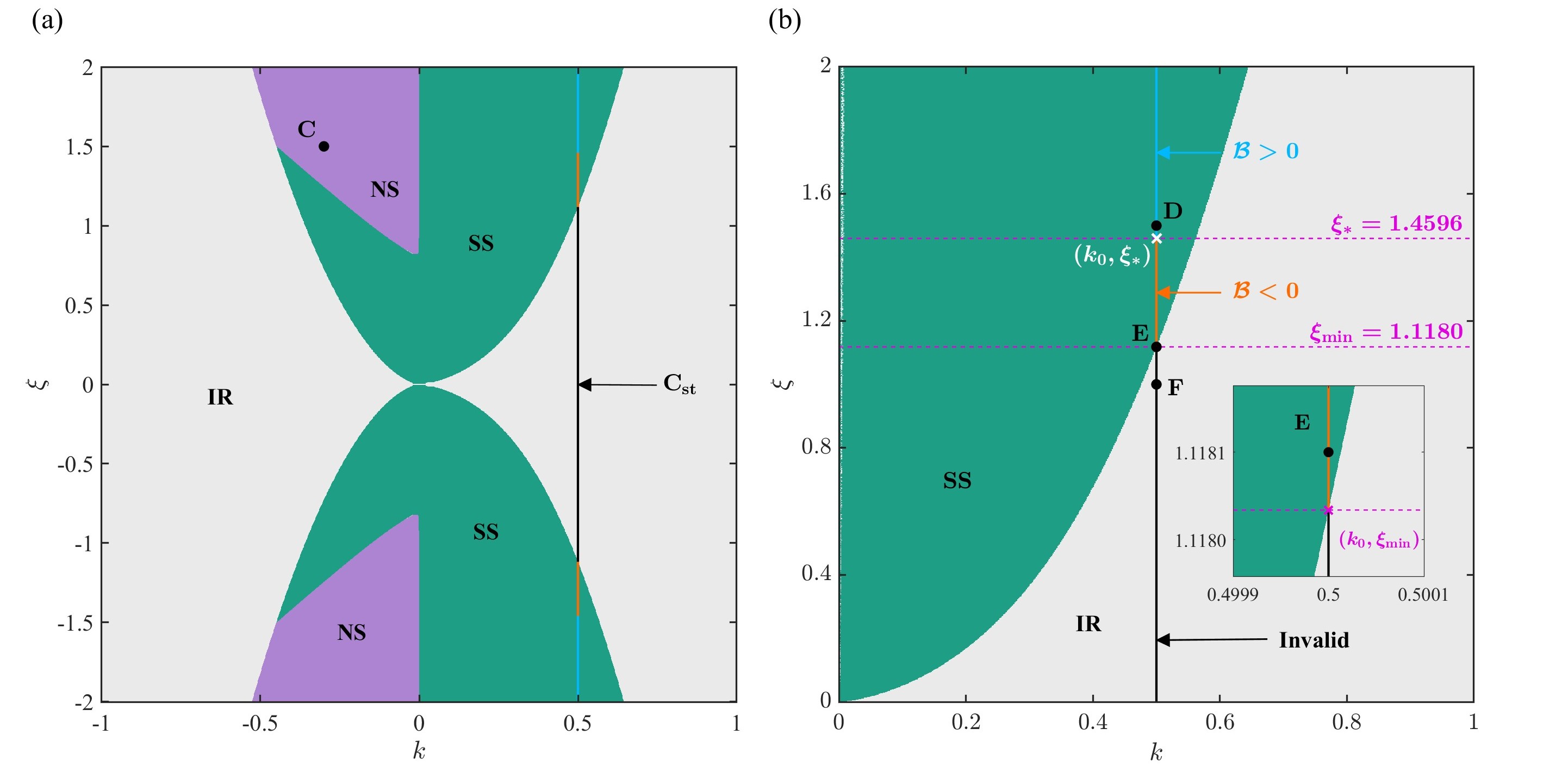}\vspace{-1em}
		\caption{\label{fig:kxi_map}
			Nondegenerate-sector morphology map for \(A=5,B=1\):
			(a) global view and (b) first-quadrant enlargement.
			\(C_{\rm st}\) contains an inactive segment in \(IR\) and
			two- and single-soliton transition segments. Points \(C,D,E,F\)
			give Figs.~\ref{fig:nondegenerate_profile},
			\ref{fig:ktrans}(a), \ref{fig:ktrans}(b) and
			\ref{fig:general_breather_inactive}, respectively;
			\(C_{\rm scd1}\) bounds the sector. Two-letter labels carry the same meaning as those in Fig.~\ref{fig:existence}.}\vspace{-1em}
	\end{figure}

\begin{figure}[h]
		\centering
		\includegraphics[width=0.4\textwidth]{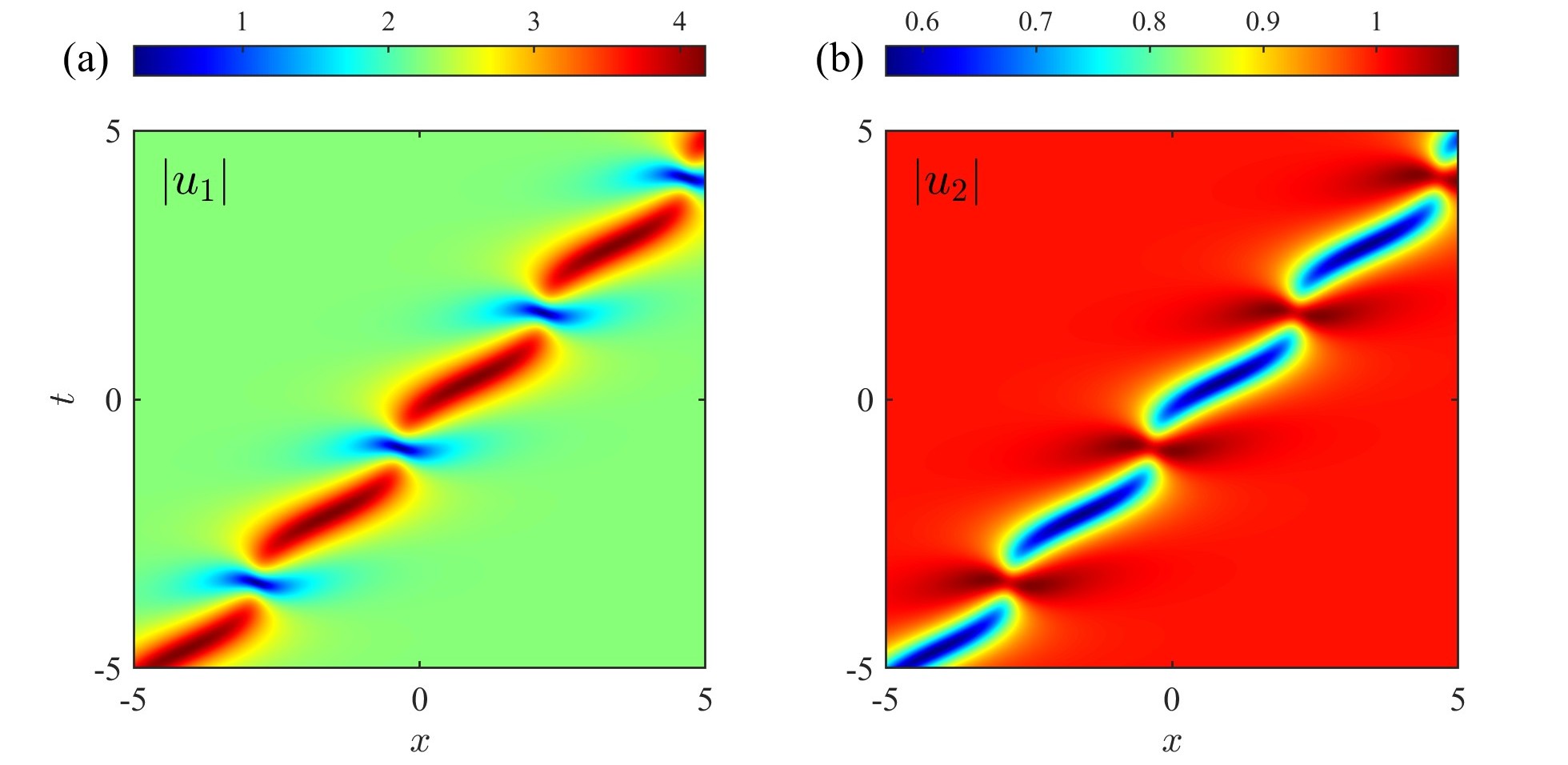}\vspace{-1em}
		\caption{Dark--four-lobed KMB at \(C\) in
			Fig.~\ref{fig:kxi_map}: (a)
			\(\lvert u_1^{[1]}\rvert\), dark; (b)
			\(\lvert u_2^{[1]}\rvert\), four-lobed. It is absent at equal
			amplitudes.}\vspace{-2em}
		\label{fig:nondegenerate_profile}
	\end{figure}

	\paragraph{Nondegenerate KMBs: State Transition induced by Frequency Match and Self-Conjugate Degeneration.---}
	\textbf{Fig.~\ref{fig:kxi_map}} shows the nondegenerate region in which the KMBs
	obtained from \(\chi_a=\chi_3\) or \(\chi_4\) are admissible. Its point \(C\) gives the dark--four-petal KMB in
	\textbf{Fig.~\ref{fig:nondegenerate_profile}}, absent at equal amplitudes. More importantly, the nondegenerate region contains a state-transition line
	\(C_{\rm st}\) for the \(\{\chi_3,\chi_4\}\) solutions, whereas the
	corresponding \(\{\chi_1,\chi_2\}\) solutions remain normal KMBs. On this
	line, the KMB is converted directly into a single- or two-soliton state by the
	one-fold Darboux transformation, without invoking any limiting procedure. The state-transition line \(C_{\rm st}\) is selected by the
	frequency-matching condition
	\begin{equation}
		\label{eq:phase_matching}
		2+k(B-A)=2-kR=0,
	\end{equation}
	which gives \(\omega_1=\omega_2\). Defining \(q=k(Ak-1)\), one equivalently has
	\(A=(q+k)/k^2\) and \(B=(q-k)/k^2\).
	At the same wavenumber, \(\Omega(\chi)=S/2\) for every spectral root, so
	\(\Omega_a=\Omega_b\) and makes the two Darboux modes share the same temporal frequency,
	thereby eliminating their relative temporal beating. The KMB intensity envelope therefore becomes a
	stationary profile in \(X=x-x_0\), where \(x_0\) denotes an arbitrary spatial translation. Its active interval begins at
	\(\xi_{\min}=2\sqrt{q^2-k^2}\). For \(\xi>\xi_{\min}\), let
	\(\eta_0=\sqrt{\xi^2-4(q^2-k^2)}\) and
	\(Y_\pm=(\eta_0\pm\xi)/2\), so that
	\((\chi_a,\chi_b)=(q+{\rm i}Y_-,q+{\rm i}Y_+)\).
	Then \(p=e^{-\xi X}>0\), and the Darboux solution reduces to
	\begin{equation}
		\label{eq:explicit_one_variable}
		\begin{aligned}
			u_j^{[1]}&=u_{j,0}Q_j(p),\\
			Q_j(p)&=1+\Delta_L
			\frac{h_{j,-}+(h_{j,-}+h_{j,+})p+h_{j,+}p^2}
			{d_-+2Lp+d_+p^2},
		\end{aligned}
	\end{equation}
	where \(h_{j,\pm}=h_j(q+{\rm i}Y_\pm)\),
	\(L=\Lambda(q+{\rm i}Y_-)=\Lambda(q+{\rm i}Y_+)\), and
	\(\Delta_L=L^*-L\). The finite coefficients \(d_\pm\) are determined by
	\(q,k,\xi\), and \(Y_\pm\). Since \(|Q_j(0)|=|Q_j(\infty)|=1\), the profile
	returns to the same background at both spatial infinities. Moreover,
	\begin{equation}
		\label{eq:complementarity}
		|u_1|^2+|u_2|^2=A+B,
		\qquad A I_1(p)+B I_2(p)=0,
	\end{equation}
	where \(I_j=|Q_j|^2-1\). Thus the localized structure arises entirely from a complementary
	redistribution of intensity: a local increase in one component is exactly
	balanced by a decrease in the other, leaving the total intensity unchanged.
	This balance identifies the resulting solitons as dark--antidark states.
	
	\begin{figure*}[t]
		\centering\vspace{-1em}
		\includegraphics[width=0.8\textwidth]{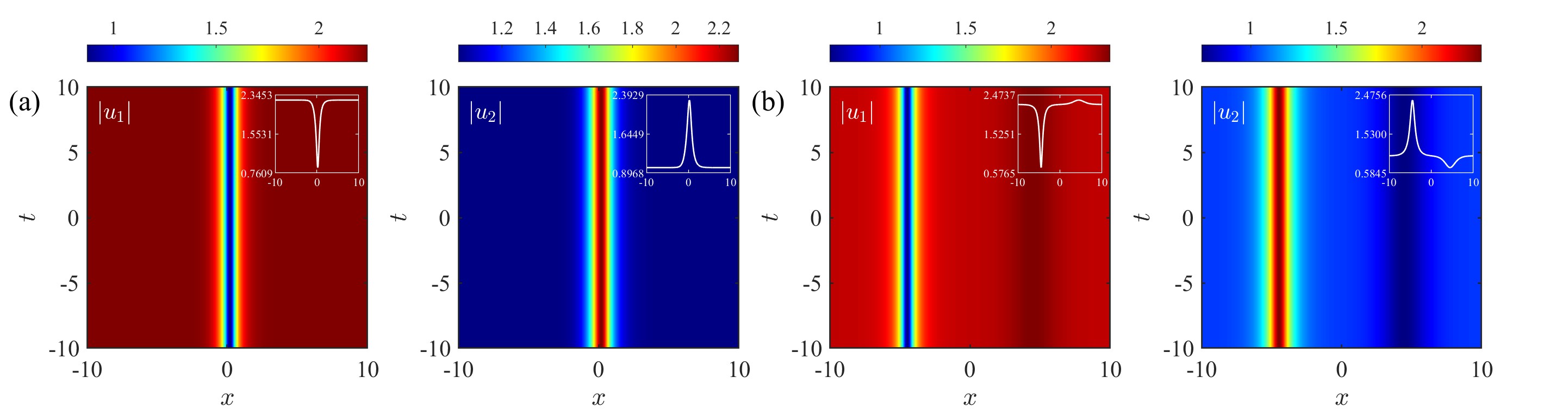}\vspace{-1em}
		\caption{Frequency-matched transitions on \(C_{\rm st}\):
			(a) single-soliton state at \(D\); (b) two-soliton state at \(E\).
			Both component moduli are shown.}\vspace{-2em}
		\label{fig:ktrans}
	\end{figure*}
	
	An exact factorization of \(I_j\) shows that the morphology is controlled entirely by
	\begin{equation}
		\label{eq:B_criterion}
		\mathcal B=(3q^2-k^2)\xi^2-4(2q^2-k^2)^2.
	\end{equation}
	The right
	tail always has a dark--antidark polarity, whereas the sign of \(\mathcal B\)
	selects the polarity of the left tail. For \(\mathcal B>0\), both tails have
	the same polarity and \(I_1<0\), \(I_2>0\) throughout \(0<p<\infty\);
	the absence of a background crossing gives a single dark--antidark soliton.
	For \(\mathcal B<0\), the left-tail polarity is reversed. Continuity then
	forces a background crossing at \(p=p_0\), and the exact factorization ensures
	that this crossing is unique. For \(\mathcal B<0\), the signs of \(I_1\) and \(I_2\) reverse once,
	producing an additional complementary hump--dip pair in which the dark and
	antidark roles of the two components are exchanged. The resulting stationary
	state therefore has a two-soliton profile. At
	\(\mathcal B=0\), the leading \(p^{-1}\) contrast vanishes and
	\(I_1\sim-C_0p^{-2}<0\), with \(C_0>0\), marking the morphology threshold.
	This threshold is
	\(\xi_{\mathcal B}=2(2q^2-k^2)/\sqrt{3q^2-k^2}\), and
	\(\xi_{\mathcal B}^2-\xi_{\min}^2
	=4q^4/(3q^2-k^2)>0\). Thus KMB activation at \(\xi_{\min}\) precedes the
	morphology transition: the interval
	\(\xi_{\min}<\xi<\xi_{\mathcal B}\) supports the two-soliton state, whereas
	\(\xi\geq\xi_{\mathcal B}\) yields the single-soliton state.
	
	\textbf{Figure~\ref{fig:ktrans}} displays the single- and two-soliton states at
	\(D\) and \(E\), respectively. Near the morphology threshold, let
	\(K_*,C_*>0\) denote constants fixed by \(q\) and \(k\) at
	\(\xi=\xi_{\mathcal B}\). As \(\mathcal B\to0^-\), the background crossing
	and the peak of the additional hump-dip pair satisfy
	\begin{equation}
		\label{eq:critical_law}
		\begin{aligned}
			&X_{\rm cross}
			=-\frac{1}{\xi_{\mathcal B}}
			\ln\frac{K_*}{|\mathcal B|}+o(1),\quad
			X_{\rm pk}
			=-\frac{1}{\xi_{\mathcal B}}
			\ln\frac{2K_*}{|\mathcal B|}+o(1),\\
			&I_{1,\rm pk}
			=C_*|\mathcal B|^2+o(|\mathcal B|^2).
		\end{aligned}
	\end{equation}
	The first two relations show that the peak and background crossing of the
	secondary soliton move together toward \(X=-\infty\), while their distance
	approaches the finite value \(\ln2/\xi_{\mathcal B}\). The shape-collapse
	analysis further confirms that this soliton retains its form and finite width,
	whereas the third relation shows that its intensity contrast vanishes
	quadratically. The two-soliton state therefore becomes a single soliton
	through the simultaneous escape and fading of the secondary soliton, rather
	than through broadening or merger with the main one. Crossing to
	\(\mathcal B>0\) does not reverse this process: the intensity deviations then
	have fixed signs, preventing both a background crossing and the formation of
	a secondary soliton. A distinct mechanism occurs at the activation threshold
	\(\xi\downarrow\xi_{\min}\). As the nonreal spectral scale vanishes, the two
	associated exponential contributions move toward opposite spatial
	infinities, and their separation grows logarithmically. Thus \(\xi_{\min}\) marks KMB activation, whereas
	\(\xi_{\mathcal B}\) marks the two-to-single-soliton morphology transition. Both are caused by an increasing separation of spatial scales.\vspace{-1em}
	
	\begin{figure}[h]
		\centering
		\includegraphics[width=0.5\textwidth]{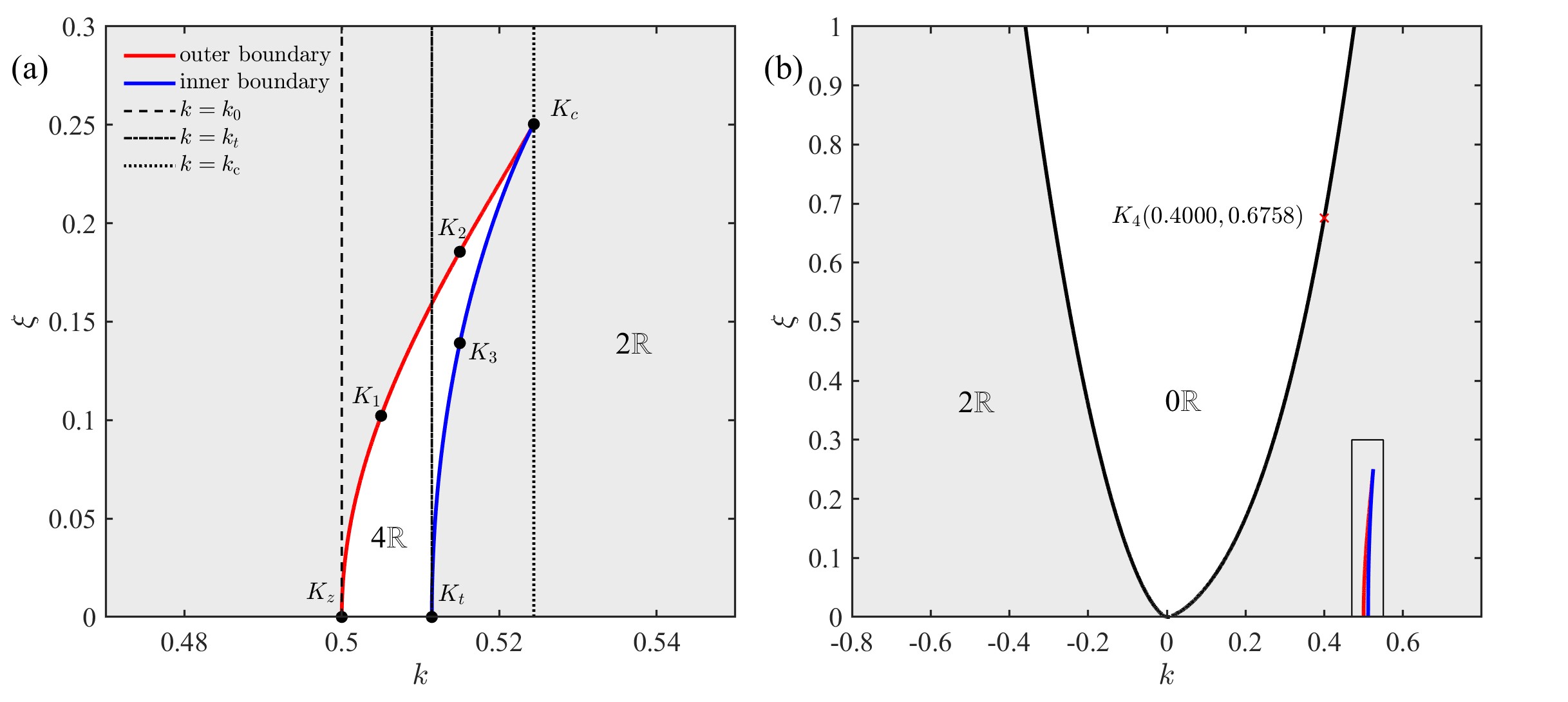}\vspace{-1em}
		\caption{Self-conjugate curves and real-root partition:
			(a) enlargement containing \(C_{\rm scd2}\) and \(C_{\rm scd3}\);
			(b) \(C_{\rm scd1}\) and point \(K_4\).
			Labels \(0\mathbb{R},2\mathbb{R},4\mathbb{R}\) give the number of real roots of the
			\(\chi\)-quartic.}\vspace{-1em}
		\label{fig:self_boundary_curve}
	\end{figure}
	
\begin{figure}[h]
		\centering\vspace{-1em}
		\includegraphics[width=0.45\textwidth]{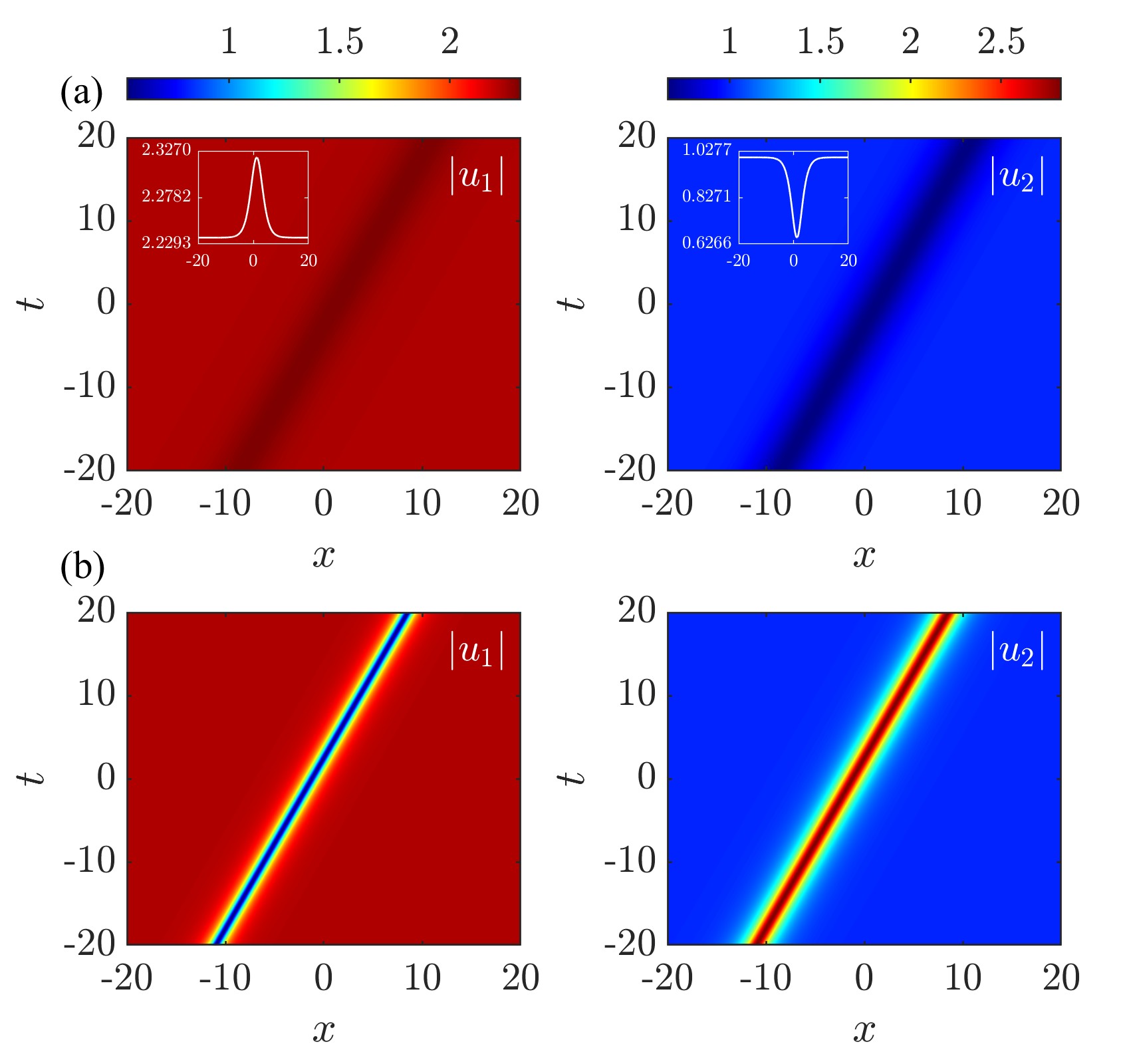}\vspace{-1em}
		\caption{Self-conjugate solitons at \(K_4\) as
			\(\xi\to\xi_{\rm sc}\). The rows correspond to the distinct
			\(\chi_{ab}\) and \(\chi_{bc}\) pairings.}\vspace{-1em}
		\label{fig:self_boundary_solitons}
	\end{figure}

	The boundary of the nondegenerate region is also a degeneration curve. On \(C_{\rm scd1}\), the \(\{\chi_3,\chi_4\}\) branch becomes
	self-conjugate:
	\((\chi_a,\chi_b)\to(\zeta,\zeta^*)\), where
	\(\zeta=\alpha-{\rm i}\xi_{\rm sc}/2\) and
	\(\Lambda(\zeta)=\Lambda(\zeta^*)=\mu\in\mathbb R\).
	\textbf{Fig.~\ref{fig:self_boundary_curve}} also shows the other two self-conjugate curves
	\(C_{\rm scd2}\) and \(C_{\rm scd3}\). With
	\(X_{\rm sc}=\xi_{\rm sc}^2\), eliminating \(\mu\) gives the curve
	\begin{equation}
		\label{eq:G_self}
		\begin{aligned}
			&G(\alpha;k,X_{\rm sc})={}
			16\alpha^4-16Sk^2\alpha^3
			+(24Rk^3-32k^2+8X_{\rm sc})\alpha^2\\
			&-4Sk^2X_{\rm sc}\alpha+X_{\rm sc}^2
			+(8k^2-2Rk^3)X_{\rm sc}
			+16k^4-8Rk^5.
		\end{aligned}
	\end{equation}
	The three self-conjugate curves are the nonnegative-real branches of
	\(\operatorname{Disc}_{\alpha}
	G=0\). For \(A=5,B=1\),
	\(C_{\rm scd1}\) is the largest-\(X_{\rm sc}\) branch and separates the
	\(2\mathbb{R}\) and \(0\mathbb{R}\) regions, whereas \(C_{\rm scd2}\) and
	\(C_{\rm scd3}\) are the two local branches enclosing the \(4\mathbb{R}\) region. Because \(L=\mu\) is real on this curve, the one-fold Darboux
	numerator proportional to \(L^*-L\) vanishes, and the unscaled solution
	reduces to the background plane wave. A finite localized state nevertheless
	survives when the curve is approached through
	\(L=\mu+{\rm i}\epsilon_L\) with
	\(\mathcal E=O(\epsilon_L)\), which makes the Darboux numerator and
	denominator vanish at the same order. This linear balance yields
	\begin{equation}
		\label{eq:self_conjugate_limit}
		u_j^{\rm sc}=u_{j,0}\left[
		1-\frac{2{\rm i}h_j(\zeta)}
		{d_{\rm sc}+K_{\rm sc}e^{\Theta_{\rm sc}}}\right],
		\qquad j=1,2,
	\end{equation}
	where
	\(\Theta_{\rm sc}=2\operatorname{Im}\zeta\,x
	+2\operatorname{Im}\Omega(\zeta)t+\ln\rho_{\rm sc}\).
	The coefficients \(d_{\rm sc}\) and \(K_{\rm sc}\) are fixed by
	\(A,B,k,\zeta\), and \(\Lambda'(\zeta)\). As the self-conjugate boundary is approached, the temporal oscillation
	frequency vanishes and the KMB period diverges. A single KMB consequently
	separates into two parallel vector solitons whose distance grows without
	bound: each component contains two mutually receding soliton traces, and the
	corresponding traces in the two components form dark--antidark pairs. The
	\(\chi_{ab}\) and \(\chi_{bc}\) root pairings characterize these two
	asymptotic solitons, shown after recentering in
	\textbf{Fig.~\ref{fig:self_boundary_solitons}}, witch are precisely the dark--antidark soliton family
	reported by Ling \emph{et al}.~\cite{LingFengZhu2018}. This identifies that family
	as the self-conjugate degeneration limit of KMBs. At the state-transition
	wavenumber \(k_{\rm st}=2/(A-B)\), their group velocity vanishes and the
	solitons are stationary; otherwise, their nonzero group velocity produces
	inclined soliton trajectories in the \((x,t)\) plane. This continuous separation within a single KMB differs
	from the non-self-conjugate transition, where approaching the degeneration
	curve from opposite sides produces two different one-sided limiting states.
	The two mechanisms also require different balances:
	\(\mathcal E=O(\epsilon_L)\) in the self-conjugate case and
	\(\mathcal E=O(\epsilon_L^{1/2})\) in the non-self-conjugate case.
	
\begin{figure}[h]
		\centering\vspace{-1em}
		\includegraphics[width=0.4\textwidth]{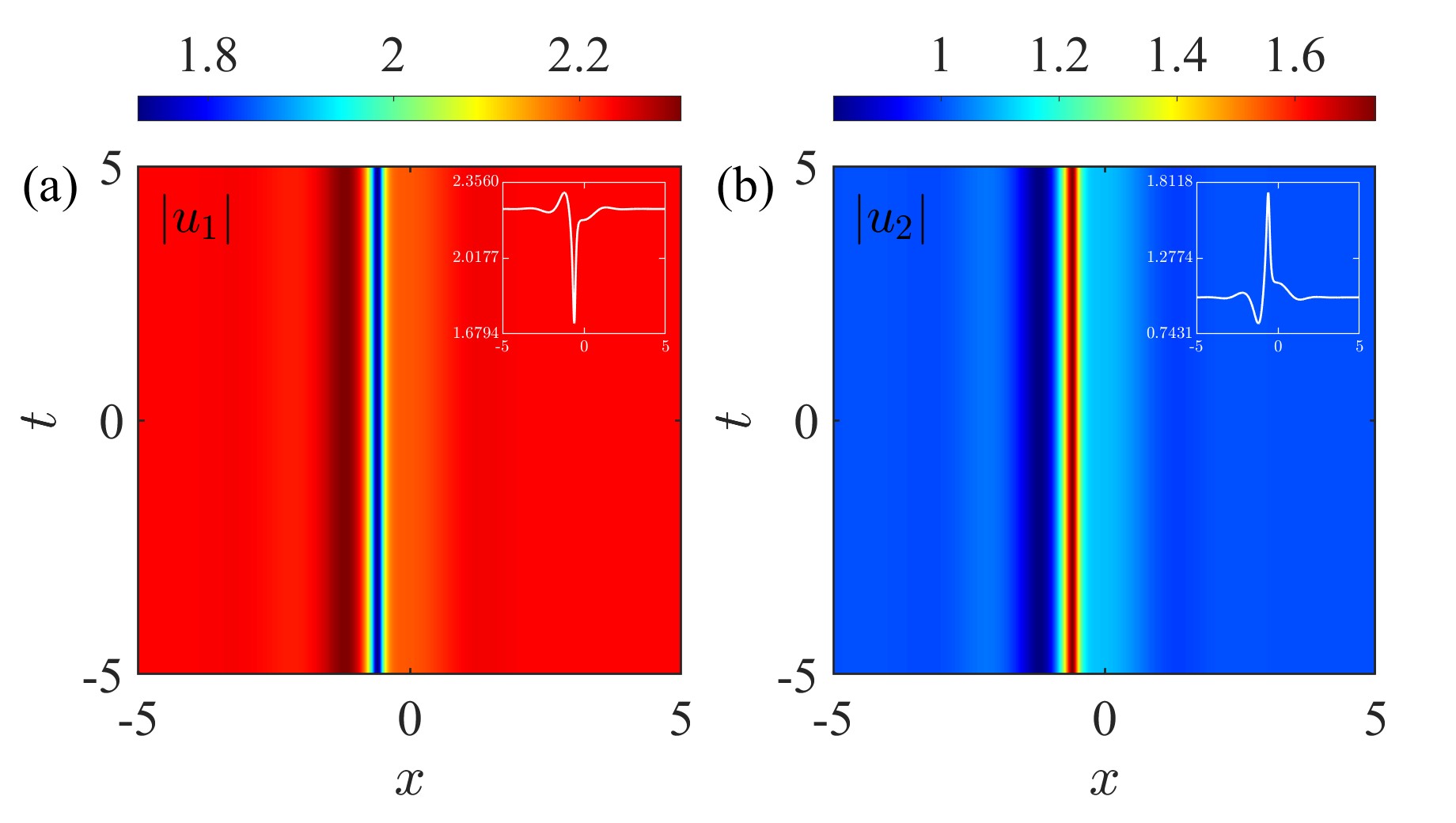}\vspace{-1em}
		\caption{General breather at \((0.5,3+1.5\mathrm{i})\) on the inactive segment of
			\(C_{\rm st}\), activated by \(\operatorname{Re}\gamma\neq0\).}
		\label{fig:general_breather_valid}\vspace{-1em}
	\end{figure}

\begin{figure}[h]
		\centering\vspace{-1em}
		\includegraphics[width=0.45\textwidth]{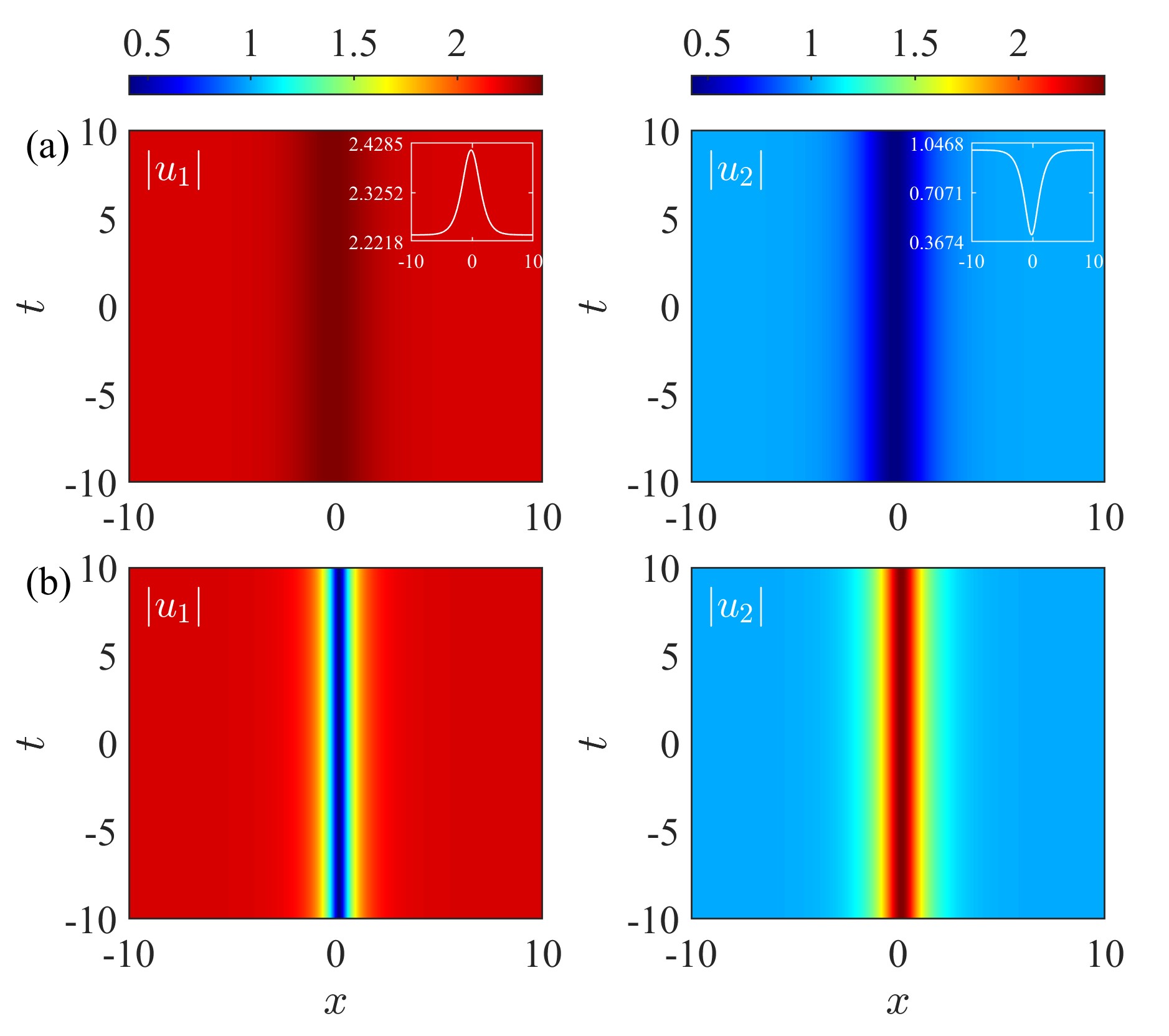}\vspace{-1em}
		\caption{KMB-line limit at \(F\), showing the logarithmic separation of
			the two activated localized structures.}\vspace{-1em}
		\label{fig:general_breather_inactive}
	\end{figure}

	\paragraph{General-breather activation and four singular limits.---}
	Fix \(k\) at the transition value and perturb the KMB modulation as
	\(\gamma=\epsilon+{\rm i}\xi\), with \(\epsilon\in\mathbb R\). The real part
	\(\epsilon\) adds a spatial oscillation to the localized exponential
	\(p=e^{(-\xi+{\rm i}\epsilon)X}\); other fixed modal phases only translate the
	carrier fringes. On the active segment of \(C_{\rm st}\), increasing
	\(|\epsilon|\) continuously converts the single-peak state into a multipeak
	general breather. On its inactive segment, the KMB correction is trivial at
	\(\epsilon=0\), but every \(\epsilon\neq0\) activates a nontrivial breather.
	Points \(G\) and \(H\) give the representative general-breather and KMB-line
	limits in \textbf{Figs.~\ref{fig:general_breather_valid}} and
	\textbf{\ref{fig:general_breather_inactive}}, respectively.
	
	The activation mechanism can be resolved analytically. Let
	\(S_0=q^2-k^2>0\), \(0<\xi^2<4S_0\), and
	\(R_0=\sqrt{4S_0-\xi^2}\). At \(\epsilon=0\), the two branches
	\(s=\pm1\) have
	\(\chi_{a,0}^{(s)}=c_s-{\rm i}\xi/2\) and
	\(\chi_{b,0}^{(s)}=c_s+{\rm i}\xi/2\), where
	\(c_s=q+sR_0/2\). Both roots share the real spectral value
	\(L_{s,0}=1/c_s\), so \(L_s^*-L_s=0\) and the Darboux correction vanishes.
	For small \(\epsilon>0\), \(L_s^*-L_s=O(\epsilon)\); hence the correction
	still disappears at every fixed \(X\). Far from the center, the two balances occur at
	\(|p|=\epsilon r\) (\(X\to+\infty\)) and
	\(|p|=r/\epsilon\) (\(X\to-\infty\)), with \(r>0\).
	Since \(e^{{\rm i}\epsilon X}=1+o(1)\) on both scales, they yield
	four recentered limits:
	\begin{equation}
		\label{eq:four_singular_limits}
		Q_{j,R}^{(s)}
		=1+\frac{\Delta_{s,1}h_{ja,0}^{(s)}}
		{d_{a,1}^{(s)}+m_0^{(s)}r},\qquad
		Q_{j,L}^{(s)}
		=1+\frac{\Delta_{s,1}h_{jb,0}^{(s)}r}
		{m_0^{(s)}+d_{b,1}^{(s)}r}.
	\end{equation}
	with \(j=1,2.\) Here \(u_{j,\ell}^{(s)}=u_{j,0}Q_{j,\ell}^{(s)}\),
	\(\Delta_{s,1}={\rm i}s\xi/(R_0c_s^2)\), and
	\(h_{j\nu,0}^{(s)}=h_j(\chi_{\nu,0}^{(s)})\).
	The finite coefficients satisfy
	\(m_0^{(s)}>0\) and
	\(d_{b,1}^{(s)}=-\overline{d_{a,1}^{(s)}}\).
	
	Each of the four limits is a regular localized solution of
	Eq.~\eqref{eq:reduced_cfl}, has a unique intensity extremum, and inherits the
	pointwise complementarity law
	\(A I_{1,\ell}^{(s)}+B I_{2,\ell}^{(s)}=0\).
	The right and left extrema occur at
	\(r_{R,*}=|d_{a,1}^{(s)}|/m_0^{(s)}\) and
	\(r_{L,*}=m_0^{(s)}/|d_{a,1}^{(s)}|\), respectively. Their spatial separation
	obeys
	\begin{equation}
		\label{eq:log_separation}
		\Delta X
		=\frac{2}{\xi}\ln\frac{1}{\epsilon}
		+\frac{2}{\xi}
		\ln\frac{m_0^{(s)}}{|d_{a,1}^{(s)}|}
		+o(1),
		\qquad \epsilon\to0^+ .
	\end{equation}
	Thus the four finite localized waves arise from a nonuniform limit: the correction
	vanishes at every fixed \(X\), while its localized parts move toward opposite
	spatial infinities with logarithmically increasing separation. The
	\(\epsilon\to0^-\) continuation reverses the spectral orientation and cannot
	be obtained by simply replacing \(\epsilon\) with \(|\epsilon|\).
	
	\paragraph{Numerical verification.---}
	To propagate the analytical states in a finite periodic domain, we isolate
	the localized KMB modulation from its plane-wave background by defining
	\(\delta u_j(x)=u_{j,{\rm KMB}}(x,0)-u_{j,0}(x,0)\) and take
	\begin{equation}
		\label{eq:tapered_initial_data}
		u_j(x,0)=u_{j,0}(x,0)+W(x)\delta u_j(x),
	\end{equation}
	where \(W(x)=1\) throughout the observation region and smoothly decreases
	to zero near the numerical boundaries. This construction preserves the full
	amplitude and phase modulation of the analytical solution in the central
	region while restoring the exact plane-wave background at the edges, thereby
	suppressing artificial boundary discontinuities and the associated
	high-frequency radiation. Physically, it launches a finite realization of
	the nonlinear modulation on the same continuous-wave background rather than
	an isolated pulse. Direct propagation reproduces the periodic localization
	and relaxation of ordinary KMBs. On the state-transition line, the suppressed
	temporal beating and the resulting single- or two-soliton intensity profiles
	are maintained, confirming that the conversion is generated by the intrinsic
	CFL dynamics rather than by an algebraic reduction of the Darboux formula.
	The limiting solitons likewise preserve their predicted velocities and
	dark--antidark structures. Only weak radiation generated in the tapered
	regions is observed, and it propagates away from the localized core without
	altering the dominant dynamics over the simulated interval.
	
	\begin{figure}
		\centering
		\includegraphics[width=0.4\textwidth]{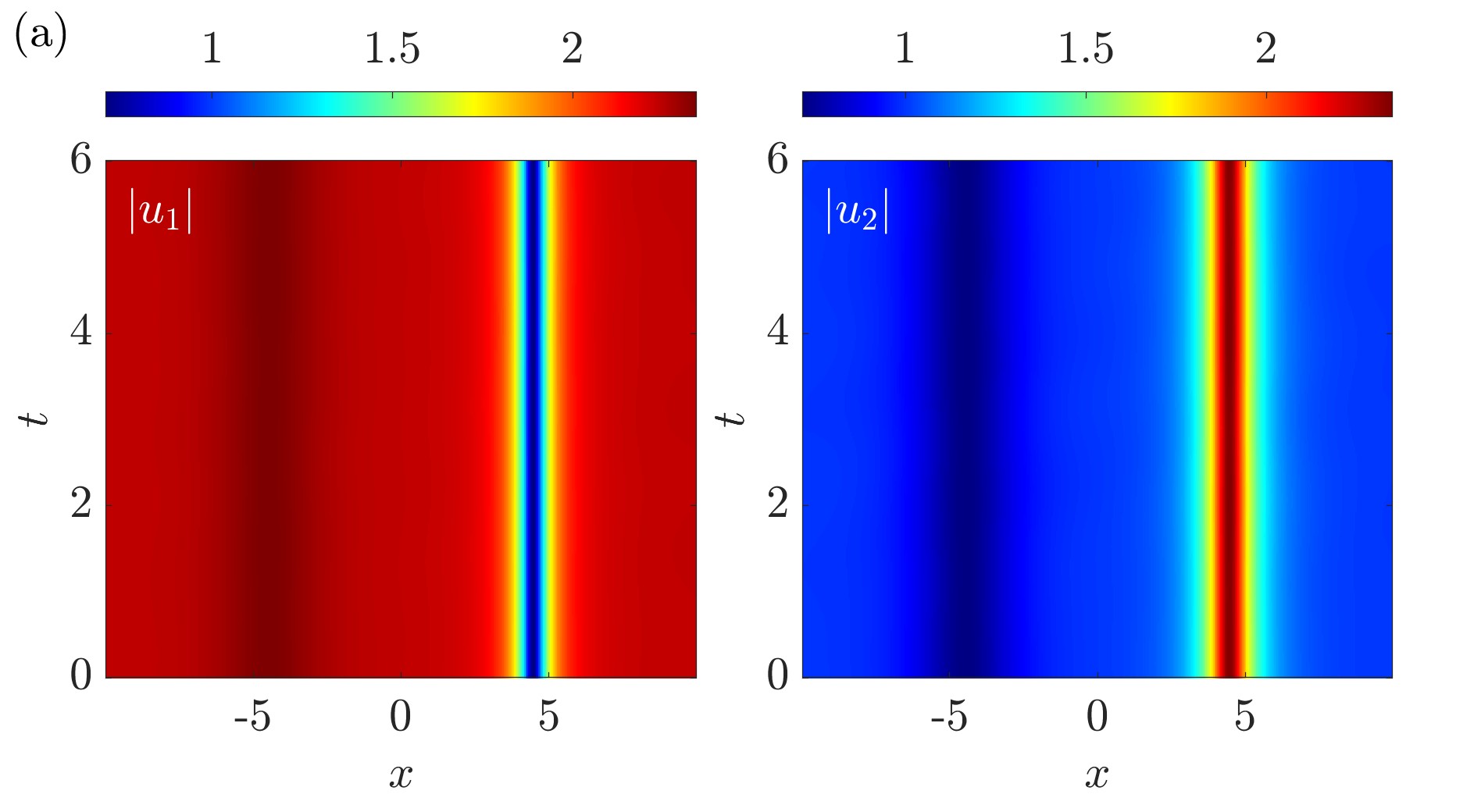}\vspace{-1.8em}
		\includegraphics[width=0.4\textwidth]{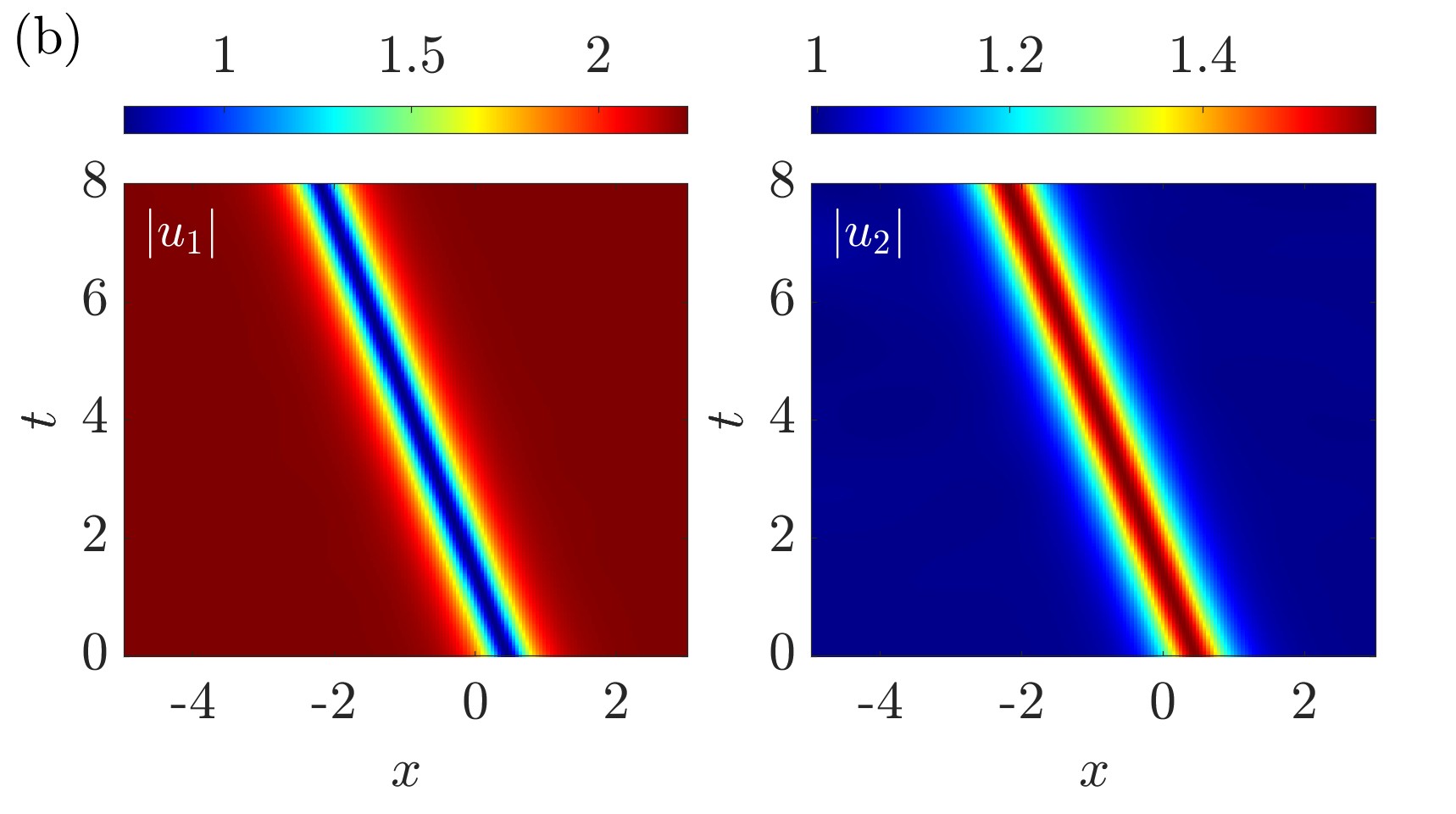}\vspace{-1em}
		\caption{Numerical propagation for \(A=5\) and \(B=1\):
			(a) state-transition solution at
			\((k,\gamma)=(0.5,1.1181{\rm i})\);
			(b) non-self-conjugate limit at
			\((k,\gamma)=(1,1.6583{\rm i})\).}
		\label{fig:num1}\vspace{-2em}
	\end{figure}
	
	\paragraph{Conclusion.---}
	Within the CFL framework, we have shown that unequal backgrounds break spectral reflection symmetry,
	enrich the vector-KMB morphologies, and enable two different types of KMB-to-soliton
	transitions. On the curve \(C_{\rm nscd}\) in the degenerate region, approaching from opposite sides produces
	dark--antidark and antidark--dark limits, yielding a discontinuous transition
	through re-pairing of \(\{\chi_a, \chi_b\}\). In the nondegenerate region, a special
	wavenumber matches the background frequencies and suppresses temporal
	beating, directly converting a one-fold KMB into a single- or two-soliton
	dark--antidark state. The complementary-intensity law and the exact
	\(\mathcal B\) criterion show that the secondary soliton disappears through
	logarithmic escape and quadratic fading rather than merger. On the
	self-conjugate boundary, the diverging KMB period separates the solution into
	two parallel dark--antidark solitons, identifying the family reported by \cite{LingFengZhu2018} as a KMB degeneration limit. A real modulation component further
	activates four localized waves in the otherwise inactive spectral interval.
	Numerical propagation supports these transition mechanisms and limiting
	states.\vspace{-2em}
	
	\section*{Acknowledgment}
	The authors acknowledge financial support from the NSFC (No.~12375002).\vspace{-1em}
	
	\bibliography{aps_nea}
	
\end{document}